\documentclass[twocolumn,10pt,aps,amsmath,amssymb,pra]{revtex4-1}
\usepackage[usenames,dvipsnames]{color} 
\usepackage{graphicx}
\definecolor{LinkColor}{rgb}{0,0,.5}
\usepackage[colorlinks=true,linktoc=BrickRed,linkcolor=BrickRed,citecolor=LinkColor,urlcolor=LinkColor]{hyperref}
\usepackage{wrapfig}

\newcommand\Id{\leavevmode\hbox{\small1\normalsize\kern-.33em1}}
\newcommand{\half}{\frac{1}{2}}
\newcommand{\ku}{\vert{0}\rangle}
\newcommand{\kd}{\vert{1}\rangle}
\newcommand{\ham}{{\mathcal{H}}}
\newcommand{\sz}{\sigma_z}
\newcommand{\ket}[1]{\left\vert{#1}\right\rangle}
\newcommand{\bra}[1]{\left\langle{#1}\right\vert}
\newcommand{\ave}[1]{\left\langle #1\right\rangle}
\newcommand{\eps}{\epsilon}
\newcommand{\carb}{$^{13}$C }
\newcommand{\kpsi}{\vert{\psi}\rangle}
\DeclareMathSymbol{\vartheta}{\mathalpha}{letters}{"12}
\DeclareMathSymbol{\theta}{\mathalpha}{letters}{"23}
\DeclareMathSymbol{\phi}{\mathalpha}{letters}{"27}
\DeclareMathSymbol{\varphi}{\mathalpha}{letters}{"1E}

\renewcommand{\emph}{\textit}

\begin{document}
\title {Spin bath narrowing with adaptive parameter estimation}
\author{Paola Cappellaro}
\email{pcappell@mit.edu}
\affiliation{Nuclear Science and Engineering Department and Research Laboratory of Electronics, Massachusetts Institute of  Technology, Cambridge, MA 02139, USA}
\begin{abstract}
We present a measurement scheme  capable of achieving the quantum limit of parameter estimation using an adaptive strategy that minimizes  the parameter's variance at each step. The adaptive rule we propose  makes the scheme robust against errors, in particular imperfect readouts, a critical requirement to extend adaptive schemes from quantum optics to solid-state sensors. 
Thanks to recent advances in single-shot readout capabilities for electronic spins in the solid state (such as Nitrogen Vacancy centers in diamond),  this scheme  can be as well applied  to estimate the polarization of a  spin bath coupled to the sensor spin. In turns, the measurement process decreases the entropy of the spin bath resulting in longer coherence times of the sensor spin.   
\end{abstract}
\maketitle

A common strategy for estimating an unknown parameter associated with a field is to prepare a probe and let it interact with the parameter-dependent field. From the probe dynamics it is possible to derive an estimator of the  parameter. The process is  repeated many times to reduce the estimation uncertainty. A more efficient procedure takes advantage of the partial knowledge acquired in each successive measurement to change the probe-field interaction in order to optimize the uncertainty reduction at each step.
This adaptive Bayesian estimation strategy has been proposed to improve the sensitivity of parameter estimation in quantum metrology~\cite{Wiseman97}. It has been shown that adaptive estimation can achieve the Heisenberg or quantum metrology limit (QML) without the need for entangled states~\cite{Higgins07,Berry09,Said11,Nusran12,Waldherr12}. 
Here we introduce a novel adaptive scheme that attains the QML,  as manifested by various statistical metrics of the estimated parameters. 
In addition, the proposed scheme can be made robust against errors so that  the QML is achieved e.g. even for imperfect readouts, a critical requirement to extend adaptive schemes from quantum optics to solid-state sensors. 
We further present an application of the adaptive scheme to the measurement of a quantum parameter: given single-shot readout capabilities for electronic spins in the solid-state~\cite{Robledo11,Morello10,Elzerman04}, the scheme could  be used to create a narrowed state of a surrounding spin bath, thus increasing the sensor coherence. In this context, the QML scaling translates into  a  shorter time  for the narrowing process, an important feature when dealing with a finite bath relaxation time.

\begin{figure}[b]
\centering
\includegraphics[scale=0.3]{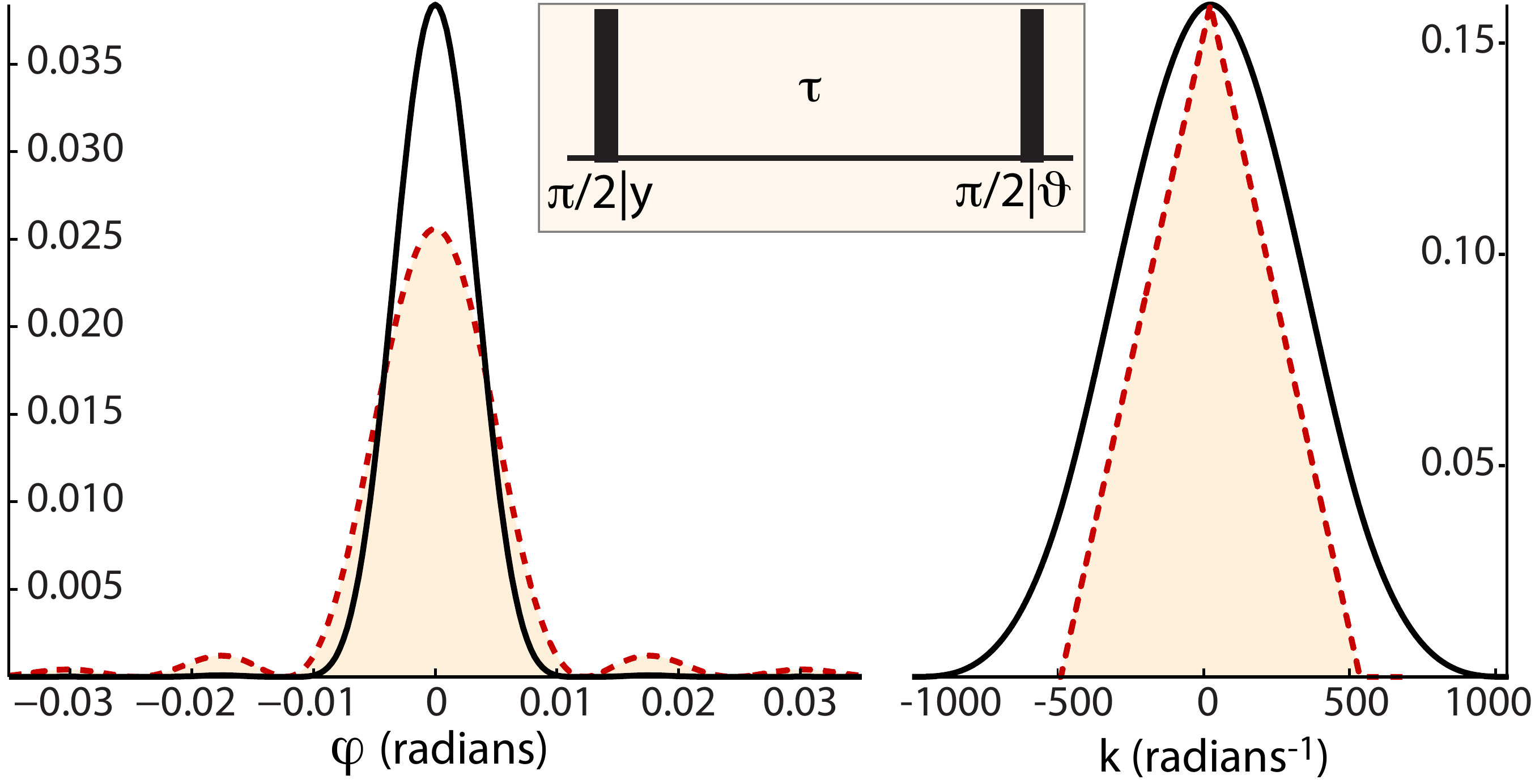}
\caption{ P.d.f (left) and its Fourier transform (right) after an 8-step adaptive measurement, with 1 (red, dashed) and 2 measurements  per step (black). In the inset, Ramsey sequence.}
\label{fig:Ramsey}\end{figure}

Consider  a two-level system $\{\ku,\kd\}$ interacting with an external field characterized by the parameter $b$, $\ham=b\sz$. A typical situation is a sensor spin-$\half$ interacting with a magnetic field. The  parameter can be estimated by a Ramsey experiment (Fig. \ref{fig:Ramsey}), where the probability of the system to  be in the  $\ket{m}$ state  ($m=\{0,1\}$) at the end of the experiment is given by
\begin{equation}
\mathcal{P}_\theta(m|b)=\half[1-(-1)^me^{-\tau/T_2}\cos(b\tau+\theta) ]
\label{eq:RamseyProb}
\end{equation}
where $\theta$ is the phase difference between the excitation and readout pulses and we introduced a decay with a constant $T_2$ during the interrogation time $\tau$. 
If we have a prior knowledge of the parameter --described by an \textit{a priori} probability distribution (p.d.f.) $P^{(0)}(b)$-- the measurement updates our knowledge, as reflected by the \textit{a posteriori} probability:
$P(b|m)\propto P^{(0)}(b)\mathcal{P}_\theta(m|b).$

More generally, after each measurement we can update the probability for the \textit{phase} $\phi=b\tau$, so that after $n$ such measurements with outcomes $\vec{m}_n$, we have a p.d.f.
\begin{equation}
P^{(n)}(\phi|\vec{m}_n)\propto P^{(n-1)}(\phi|\vec{m}_{n-1})\mathcal{P}_\theta(m_n|\phi)
\label{eq:posterior}
\end{equation}
Thanks to the periodicity of the probability $P(\phi)$, we can expand it in Fourier series~\cite{Said11},  
$P^{(n)}(\phi)=\sum_k p^{(n)}_ke^{ik\phi}$, 
so that we can rewrite Eq. (\ref{eq:posterior})  as
\[
\begin{array}{ll}
p_k^{(n)}\propto \half{p_k^{(n-1)}}&+\frac14{e^{-\tau/T_2}}\times\\&\left[e^{i(m_n\pi+\theta)}{p_{k-1}^{(n-1)}}+e^{-i(m_n\pi+\theta)}{p_{k+1}^{(n-1)}}\right]
\end{array}
\]
The proportionality factor is set by imposing that $p_0^{(n)}=\frac1{2\pi}$ as required for a normalized p.d.f. 
We can further generalize this expression when the system is let evolve for an \textit{integer} multiple $t_n$ of the time $\tau$, thus obtaining a general update rule for the p.d.f.:
\begin{equation}
\begin{array}{ll}
p_k^{(n)}\propto& \half{p_k^{(n-1)}}+\frac14{e^{-t_n\tau/T_2}}\times\\& \left[e^{i(m_n\pi+\theta_n)}{p_{k-t_n}^{(n-1)}}\right.\left.+e^{-i(m_n\pi+\theta_n)}{p_{k+t_n}^{(n-1)}}\right]
\end{array}
\label{eq:update}
\end{equation}
An adaptive strategy will then seek to choose at each step the optimal $t_n$ and $\theta_n$ that lead to the most efficient series of $N$ measurements for a desired final uncertainty. 

In order to design an adaptive strategy, we need to define a metric for the uncertainty (and accuracy) of the estimate. 
The Fourier transform of the p.d.f.  can be used to calculate the moments of the distribution as well as other metrics and estimator. 
From the formula for the moments,
$\ave{\phi^\alpha}=\int_{-\pi}^\pi P(\phi)\phi^\alpha d\phi=\sum_k p_k \int_{-\pi}^\pi e^{ik\phi}\phi^\alpha d\phi$ 
we can calculate the variance,
\[\ave{\phi^2}-\ave{\phi}^2=\frac{2 \pi ^3}{3}p_0+4\pi\sum_{k\neq0}\frac{(\text{-}1)^k}{k^2}p_k-\ave{\phi}^2, \]
where the average is
$\ave{\phi}=-2i\pi\sum_{k\neq0}\frac{(\text{-}1)^k}kp_k.$

The variance is often not the best estimate of the uncertainty for a periodic variable~\cite{Berry09}. A better metric is the Holevo variance~\cite{Holevo84},
\begin{equation}
V_H=(2\pi|\ave{e^{i\phi}}|)^{\text{-}2}-1=(2\pi|p_{\text{-}1}|)^{\text{-}2}-1,
\label{eq:holevo}
\end{equation}
where we used the fact that $\ave{e^{i\phi}}=p_{\text{-}1}$. 
We further notice that while the absolute value of $p_{\text{-}1}$ gives the phase estimate uncertainty, its argument provides an unbiased estimate of $\phi$. More generally,  estimates are given by  $\phi_{est}=\arg(\ave{e^{it\phi}})/t=\arg(p_{-t})/t$,  giving  a new meaning to the Fourier coefficients of the p.d.f.

The goal of the estimation procedure is then to make  $|p_{\text{-}1}|$  as large as possible. 
Assume for simplicity $\phi=b\tau=0$ and neglect any relaxation. Then the probability of the outcome $m_s=0$ is $\mathcal{P}_\theta(0|0)=\half(1-\cos\theta)$.   We assume that we do not have any a priori knowledge on the phase, so that $P^{(0)}(\phi)=1/2\pi$. We fix the number of measurements, $N$, each having an interrogation time $T_n=t_n\tau=2^{N-n}\tau$~\cite{Giedke06,Boixo08,Said11}. 
A potential strategy would be to maximize $|p_{-1}^{(n)}|$ at each step $n$. However, under the assumptions made, $p_{-1}^{(n)}=0$ until the last step, $n=N$, where it is
\[p^{(N)}_{-1}=\frac{e^{-i(m_N\pi+\theta_N)}}{4\pi} \left(2\pi p^{(N-1)}_{-2}e^{2i\theta_N}+1\right)\]
Writing $p^{(N-1)}_{-2}=qe^{i\chi}$, we have
\[4\pi|p^{(N)}_{-1}|=\sqrt{1+4\pi^2q^2+4\pi q\cos(\chi+2\theta_N)}\]
This is maximized for $\theta_N=-\chi/2=\half\arg(p^{(N-1)}_{-2})$ and by maximizing $q=|p^{(N-1)}_{-2}|$. 
A similar argument holds for the maximization of $|p^{(N-1)}_{-2}|$: one has to set $\theta_{N-1}=\half\arg(p^{(N-2)}_{-4})$ and maximize $|p^{(N-2)}_{-4}|$. 
By recursion we have that at each step we want to maximize
\[|p^{(n)}_{-t_n}|=\left|\frac{e^{-i(m_n\pi+\theta_n)}}{4\pi} \left(2\pi p^{(n-1)}_{-t_{n-1}}e^{2i\theta_n}+1\right)\right|\]
We have thus found a good  adaptive  rule, which fixes $t_n=2^{N-n}$ and $\theta_n=\half\arg\left(p^{(n-1)}_{-t_{n-1}}\right)$.

With this rule we obtain the standard quantum limit (SQL) for the phase sensitivity, as we now show.
Using the optimal phase,  the Fourier coefficients $p_{-t_n}^{(n)}$ are at each step
\[p_{-t_n}^{(n)}=\half\left(\frac1{2\pi}+p_{-t_{n-1}}^{(n-1)}\right)=\frac1{2\pi}(1-2^{-n})\]
Then, for a total number of measurements $N$, the Holevo variance is $V_H=(1-2^{-(N+1)})^{-2}-1\approx
2^{-N}$. The total interrogation time is $T=\tau(2^{N+1}-1)$ yielding 
\begin{equation}
V_H(T)=\frac{4 T \tau }{(T-\tau )^2}\approx\frac{4\tau}T
\label{eq:Holevo1pass}
\end{equation}

We can improve the sensitivity scaling and reach the QML by a simple modification of this adaptive scheme.  Instead of performing just one measurement of duration $t_n$ at each $n^{th}$ step, we perform two, updating the p.d.f. according to the outcomes. For $\phi=0$ the update rule at each step is now
\[
p_k^{(n)}=\frac1{\mathcal{N}}\left[6p_k^{(n-1)}+4p_{k-t_n}^{(n-1)}+4p_{k+t_n}^{(n-1)}+p_{k-2t_n}^{(n-1)}+p_{k+2t_n}^{(n-1)}\right]
\]
with the normalization factor
\[\mathcal{N}=2\pi\left[6p_0^{(n-1)}+p_{-2t_n}^{(n-1)}+p_{2t_n}^{(n-1)}\right].\]
Restricting the formula above to the terms $p^{(n)}_{-t_n}$ 
gives
\begin{equation}p^{(n)}_{-t_n}=\frac{\frac1{2\pi}+p^{(n-1)}_{-t_{n-1}}}{\pi\left(\frac3{2\pi}+p^{(n-1)}_{-t_{n-1}}\right)}\label{eq:twosteps}
\end{equation}
By recursion this yields 
\[|p^{(n)}_{-t_n}|=\frac{1}{2 \pi }\left(1-\frac{3}{2^{2n+1}+1}\right),\]
from which we obtain a Holevo variance that follows the QML, $V_H\approx3\cdot2^{-2N}$, or in terms of the total interrogation time
\begin{equation}
V_H=\frac{48 T \tau ^2 (T+4 \tau )}{(T-2 \tau )^2 (T+6 \tau )^2}\approx\frac{48\tau^2}{T^2}.
\label{eq:Holevo2pass}
\end{equation}

The classical and quantum scaling of the adaptive scheme with one or two measurements per step is confirmed by the p.d.f. obtained in the two cases (Fig. \ref{fig:Ramsey}). For one measurement, the final p.d.f Fourier coefficient  are $|p_k|=\frac1{2\pi}(1-2^{-(N+1)}|k|)$ and the probability is well approximated by a sinc function,
\[P^{(N)}(\phi)=\frac{2^{N+1}}{2\pi}\text{sinc}(2^{N+1}\phi)^2,\]
which gives a variance $\sigma\approx2^{-N/2}$.
For two measurements per step, instead, the p.d.f. is well approximated by a Gaussian (see Appendix)  with  a width $\sigma =\frac{\sqrt{3}}2 \cdot2^{-N}$.

We now consider possible sources of non-ideal behavior. The first generalization is to phases $\phi\neq0$. In this case, while the SQL is still achieved with the one-measurement scheme,  two measurements per step do not always  reach the QML. Indeed, at each step there is a probability $\mathcal{P}(1-\mathcal{P})$  that the two measurements will give different results; if this happens at the $n^{\text{th}}$ step,  we obtain $p^{(n)}_{-t_n}=0$, thus failing to properly update the p.d.f. While the probability of failure is low, a solution could be to perform three measurements and update the p.d.f. only based on the majority vote.

\begin{figure}
\centering
\includegraphics[scale=0.3]{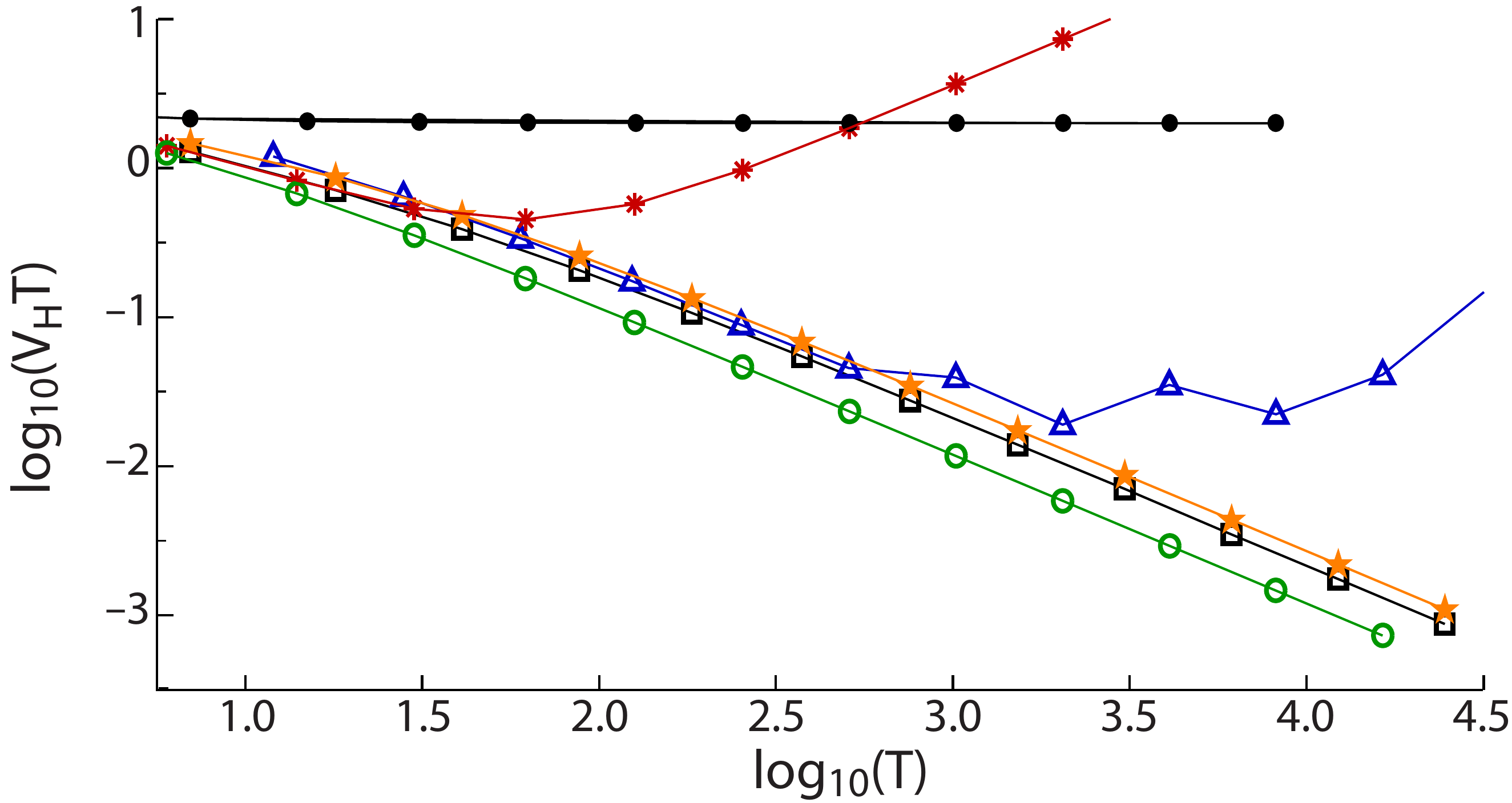}
\caption{Holevo variance vs. total time $T=M\tau(2^{N+1}-1)$, with $\tau=1$. (\large$\bullet$\normalsize) with 1 measurement per step (M=1) $V_H$ follows the SQL. (\large$\circ$\normalsize) 2 measurements per step (M=2) achieve the QML. (\large$*$\normalsize) with c=0.95, the QML scaling is lost, but can be preserved for longer with M=4  (\footnotesize$\triangle$\normalsize) and  restored (\footnotesize$\square$\normalsize) by setting M=$n+1$,  even for lower c, e.g. c=0.85 ($\star$)}
\label{fig:Holevo}
\end{figure}

We can further consider the cases where the signal decays due to relaxation or there is an imperfect readout. Then the probability (\ref{eq:RamseyProb}) becomes
\[\mathcal{P}_\theta(m|b)=\half[1-c(-1)^me^{-\tau/T_2}\cos(b\tau+\theta) ],\]
with $c$ the readout fidelity. Considering the effects of only this (constant) term, the update rule Eq. 
\ref{eq:twosteps} becomes
\[p^{(n)}_{-t_n}=\frac{c\left(\frac1{2\pi}+p^{(n-1)}_{-t_{n-1}}\right)}
{\pi\left(\frac1{\pi}\left(1+\frac{ c^2}2\right)+c^2p^{(n-1)}_{-t_{n-1}}\right)}\]
We can calculate a recursion relationship in the limit of good measurement,  $\eps=(1-c)\approx0$, to obtain
\[
|p_{-1}|\approx\frac{1}{2 \pi }\left(1-\frac{3}{2}(1+N\eps)2^{-2N}\right), 
\]
which yields an Holevo variance $V_H\approx3(1+\eps N)2^{-2N}$ that does not follow anymore the QML scaling,  except for  $\eps N\sim1$. A similar, more complex result is expected if relaxation effects are taken into account (see Appendix). A strategy to overcome this limitation is to repeat the measurement at each step more than two times (Fig. \ref{fig:Holevo}). Specifically, setting the number of measurements M=$n+1$ (if allowed by  relaxation constraints) restores the QML scaling.

The proposed adaptive method  promises to achieve Heisenberg-limited estimation of a classical phase without the need of fragile entangled states and thus it could improve the sensitivity e.g. of recently proposed magnetic sensors~\cite{Taylor08}. It can be as well used to measure a quantum variable, such as a phase resulting from the coupling of the sensor to a larger quantum system or bath. 
In turns, the measurement can be used to lower the entropy of the bath (usually a thermal equilibrium mixture)  yielding an increase in the coherence time of the sensor~\cite{Giedke06,Imamoglu03,Klauser06}. The QML scaling of this adaptive method translates into a faster narrowing of the bath dispersion, which would improve similar schemes in solid-state systems~\cite{Bluhm10,Togan11}, where the bath itself might present fluctuations.

Specifically, we consider the coupling of a sensor spin to a spin bath. This situation is encountered in many physical systems, such as quantum dots~\cite{Hanson05,Bluhm11} or phosphorus donors in silicon~\cite{Morello10,Abe10}. Here we  analyze as an example the system comprising 
  a Nitrogen-Vacancy (NV) center electronic spin coupled to the bath of nuclear \carb spins in the diamond lattice~\cite{Jelezko06,Childress06}. Recent advances in the measurement capabilities~\cite{Robledo11} offer single-shot read-out of the NV state, thus enabling adaptive schemes. 
  
In a large magnetic field along the NV axis, the hyperfine interaction between the  electronic spin and the nuclear spins is truncated to its secular part, $\ham=S_z\sum_kA_kI_{z,k}=S_zA_z$ (where $S$ denotes the electronic spin,  $I_k$ the nuclear spins). During a Ramsey sequence on resonance with the $m_s=0,1$ energy levels of the electronic spins, the coupled system  evolves as 
\begin{equation}
\ket{\psi(t)}=\left[\sin\left(A_zt\right)\kd+\cos\left(A_zt\right)\ku\right]\kpsi_C,
\label{eq:stateevolution}
\end{equation}
where $\kpsi_C$ is the initial state of the nuclear spin bath. 
The measurement scheme  (Ramsey followed by NV read-out) is a quantum non-demolition measurement\cite{Braginsky80,Mlynek97,Waldherr11} for the nuclear spins, since their observable does not evolve -- as long as the secular approximation holds. 
The adaptive process  is then equivalent to determining the state-dependent (quantized) phase $\phi=\ave{A_zt}$. The  uncertainty on the nuclear bath state, $\rho_C=\!\sum_\alpha p_\alpha \ket{\psi_\alpha}\!\!\bra{\psi_\alpha}_C$,  is reflected in the  p.d.f of the phase  (with an injective relation if the operator $A_z$ has non-degenerate eigenvalues). Thus updating the phase p.d.f. will update the density operator describing the state of the nuclear bath. After each readout of outcome $m$, the system is in the state
\begin{equation}
\begin{array}{ll}
\rho^{(n)}&\propto  \ket{m}\!\bra{m}\!\rho^{(n-1)}\!\ket{m}\!\bra{m}\\
&= \ket{m}\!\!\bra{m}\sum_\alpha \mathcal{P}_\theta(m|\phi_\alpha)p_\alpha^{(n-1)}\ket{\psi_\alpha}\!\!\bra{\psi_\alpha}_C,\end{array}
\label{eq:stateupdate}
\end{equation}
with $\mathcal{P}_\theta(m|\phi_\alpha)\!=\!\left|\bra{m,\psi_\alpha}\!\left[\sin(A_zt)\kd\!+\!\cos(A_zt)\ku\right]\!\ket{\psi_\alpha}\right|^2\!$.
Note that in this expression the probability update rule is equivalent to Eq. \ref{eq:posterior}  and thus the adaptive procedure ensures that the final state has lower entropy than the initial one.

A difference between measuring a classical field and a quantum operator is that in the latter case the resulting phase is quantized, thus it  has a discrete p.d.f.. An extreme case is when all the couplings to the $N_C$ nuclear spins are equal, $A_k=a$, $\forall k$. Then the eigenvalues are $na/2$, with $|n|\leq N_C$ integer, each with a degeneracy $d(n)=\binom{N_C}{N_C/2+n}$. While the adaptive scheme needs to be modified (e.g. by considering a discrete Fourier transform), we note that since all the eigenvalues are an integer multiple of the smallest, non-zero one ($a$ for $N_C$ even,  $a/2$ for $N_C$ odd), we only need $M$ steps, with $2^M\geq \frac {N_C}2$ [with minimum interrogation time $\tau=2\pi/(a2^M)$], to achieve a perfect measurement of the degenerate phase $\phi$~\cite{Giedke06}.
\begin{figure}[t]
\centering
\includegraphics[scale=0.3]{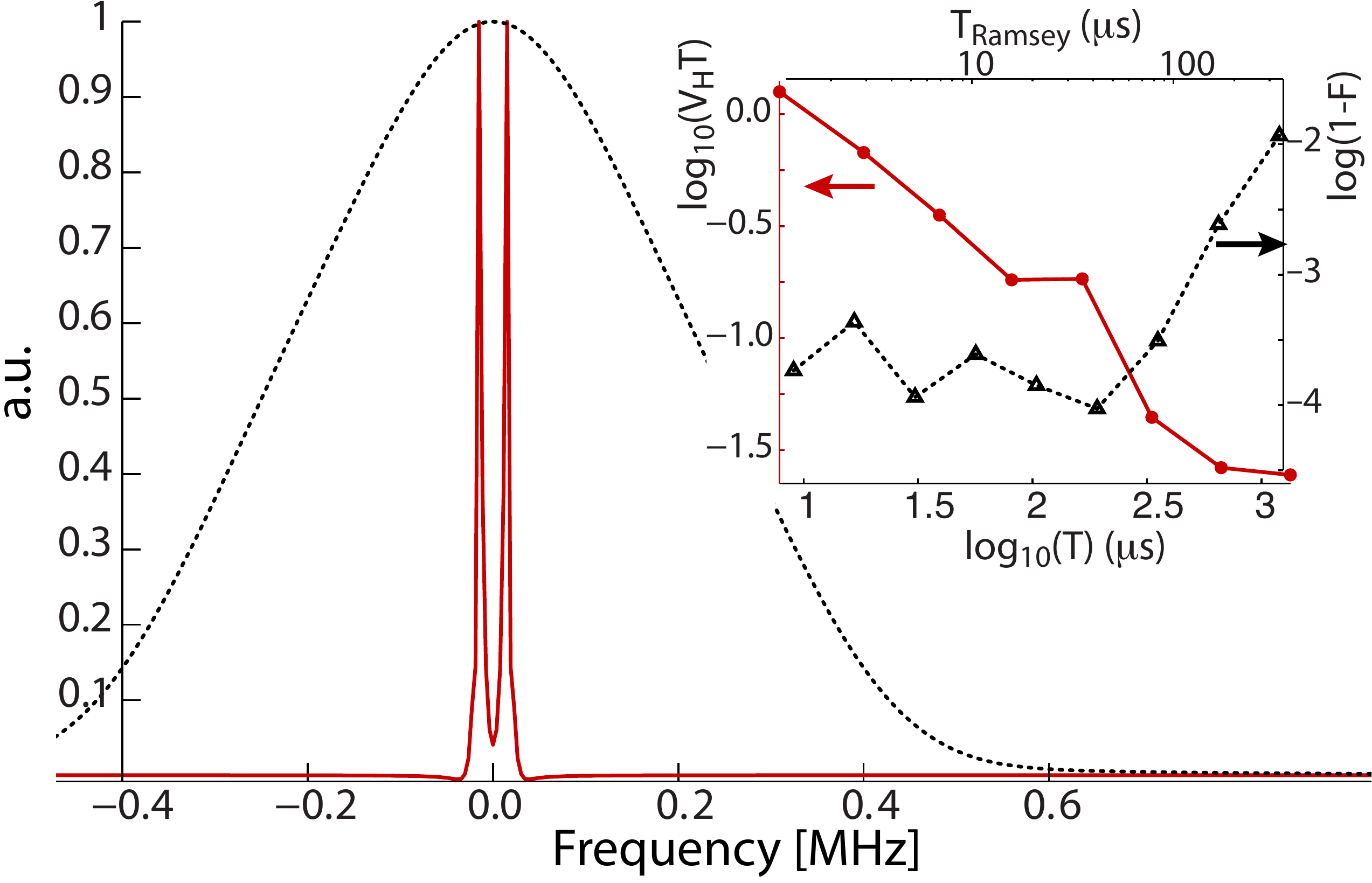}\vspace{-12pt}
\caption{Simulated NV center spectrum from a Ramsey sequence before (dotted) and after  (solid line) an 8-step adaptive measurement. 
The adaptive scheme narrows the bath distribution, here corresponding to a 50$\times$ improvement of the coherence time.
We considered $\sim2600$ closest nuclear spins, randomly positioned in the diamond lattice, in a $1$T magnetic field and initially  in the  maximally mixed-state. We simulated full dipolar couplings among the nuclear spins and between the NV and the bath spins (approximating the hyperfine interaction)  to validate the secular approximation.
The narrow spectrum reveals an average nuclear field of $\ave{B_\textrm{n}}\approx 700$nT (this was chosen at random among  possible nuclear spin state-dependent fields; similar results were obtained for different fields and  spin positions in the lattice). In the inset, Holevo variance (circles) as a function of the total time, and signal decay (dotted line) due to intra-bath couplings during the adaptive scheme interrogation time ($T_{\text{Ramsey}}$).}
\label{fig:NVsims}
\end{figure}
In the more common scenario where $A_k$ varies with the nuclear spin position (and $N_C$ is large enough) the eigenvalues give rise to an almost continuous phase~\cite{Klauser06}, thus it is possible to directly use the adaptive scheme derived above.
 
As an example of the method, we consider one NV center surrounded by a bath of nuclear spins (\carb with 1.1\% natural abundance). At low temperature and for NV with low strain, it is possible to perform single-shot readout of the electronic spin state with high fidelity in tens of $\mu$s~\cite{Robledo11}. Optical illumination usually enhances the electronic-induced nuclear relaxation~\cite{Jiang09}, due to the non-secular part of the hyperfine interaction.
 This effect  is however quenched in a high magnetic field ($B\geq 1$T)  and the relaxation time is much longer than the measurement time ($T_1\geq 3$ms~\cite{Neumann10b}), sign of a good QND measurement.

We simulated the Ramsey sequence and adaptive measurement with a bath of $\sim 2600$ spins around the spin sensor in a large magnetic field.  We considered the full anisotropic hyperfine interaction between the NV and the \carb spins and we took into account intra-bath couplings with a disjoint cluster approximation~\cite{SOM,Maze08b}.  Even for the longest evolution time of the Ramsey sequence required by the adaptive scheme, the fidelity $F$ of the signal with the ideal Ramsey oscillation (in the absence of couplings) is maintained. After an 8-step adaptive measurement, the nuclear spin bath is in a narrowed state.  
We note that in general the adaptive scheme does not polarize the spin bath (indeed a final low polarization state is more probable). However, the bath purity is increased, which is enough to ensure longer coherence times for the sensor spins, since it corresponds to a reduced variance of the phase and hence of the sensor spin dephasing.
In Fig.~\ref{fig:NVsims} we compare the NV center spectrum for an evolution under a maximally mixed nuclear spin bath and under the narrowed spin bath. The figure shows a remarkable improvement of the NV coherence time. 

In conclusion, we described an adaptive measurement scheme that has the potential to achieve the quantum metrology limit for a classical parameter estimation. We analyzed how imperfections in the measurement scheme affect the sensitivity and proposed strategies to overcome these limitations. This result could for example improve the sensitivity of spin-based magnetometers, without recurring to entangled states.
In addition, we applied the scheme to the measurement of a quantum parameter, such as arising from the coupling of the sensor to a large spin bath. We showed that the adaptive scheme can  be used to prepare the spin bath in a narrowed state: as the number of possible configurations for the spin bath is reduced, the coherence time of the sensor  is increased. The scheme could then be a promising strategy  to increase the coherence time of  qubits, without the need of dynamical decoupling schemes that  have large overheads and interfere with some magnetometry and quantum information tasks.
\vspace{24pt}

\textbf{Acknowledgments} -- This research was supported in part by the
U.S. Army Research Office through a MURI grant No. W911NF-11-1-0400.

\appendix

\onecolumngrid
\section{Probability distribution for the adaptive scheme with two measurements per step}
In the main text we considered the adaptive scheme where at each step two measurements (with the same reading time) are carried out and presented an approximate formula for the p.d.f that is obtained after $N$ steps. Here we present details of the derivation for the case where $\phi=0$. 
\\
In the Fourier space, the p.d.f. coefficients can be calculated from the recursive relation to be 

\[p_k=\left\{\begin{array}{ll}
\frac1{2\pi}\frac{\left(2^{N+2}-1-|k|\right) \left(2^{N+2}-|k|\right) \left(2^{N+2}+1-|k|\right)
-4 \left(2^{N+1}-1-|k|\right) \left(2^{N+1}-|k|\right) \left(2^{N+1}+1-|k|\right)}
{ \left(2^{N+2}+2^{3 N+5}\right)}
&k\leq2^{N+1}-2\\
\frac1{2\pi}\frac{\left(2^{N+2}-1-|k|\right) \left(2^{N+2}-|k|\right) \left(2^{N+2}+1-|k|\right)}{ \left(2^{N+2}+2^{3 N+5}\right)} 
&k>2^{N+1}-2
\end{array}\right. 
\]
For large $N$ we can simplify the expressions as:
\[p_k=\left\{\begin{array}{ll}
\frac1{2\pi}\frac{ \left(2^{N+2}-|k|\right)^3-4 \left(2^{N+1}-|k|\right)^3}
{ \left(2^{N+2}+2^{3 N+5}\right)}\approx  \frac1\pi\left[\left(1-2^{-(N+2)}|k|\right)^3-\half\left(1-2^{-(N+1)}|k|\right)^3\right]
&k\leq2^{N+1}-2\\
\frac1{2\pi}\frac{ \left(2^{N+2}-|k|\right)^3}{ \left(2^{N+2}+2^{3 N+5}\right)} \approx \frac1\pi\left(1-2^{-(N+2)}|k|\right)^3
&k>2^{N+1}-2
\end{array}\right. \]

In turn, these expressions are well approximated by a Gaussian (although the original function has longer tails),  
\[p_k\approx \frac{e^{-3 k^2 2^{-2 N-3}}}{2 \pi }\]
with Fourier transform
\[P(\phi)\approx\frac{e^{-\frac{2}{3}\left(2^N \phi \right)^2}}{2^{-N}\sqrt{{3 \pi }/{2}} }\]

\section{Influence of noise on the adaptive scheme}
\begin{wrapfigure}{r}{0.5\textwidth}
\begin{center}
\includegraphics[width=0.5\textwidth]{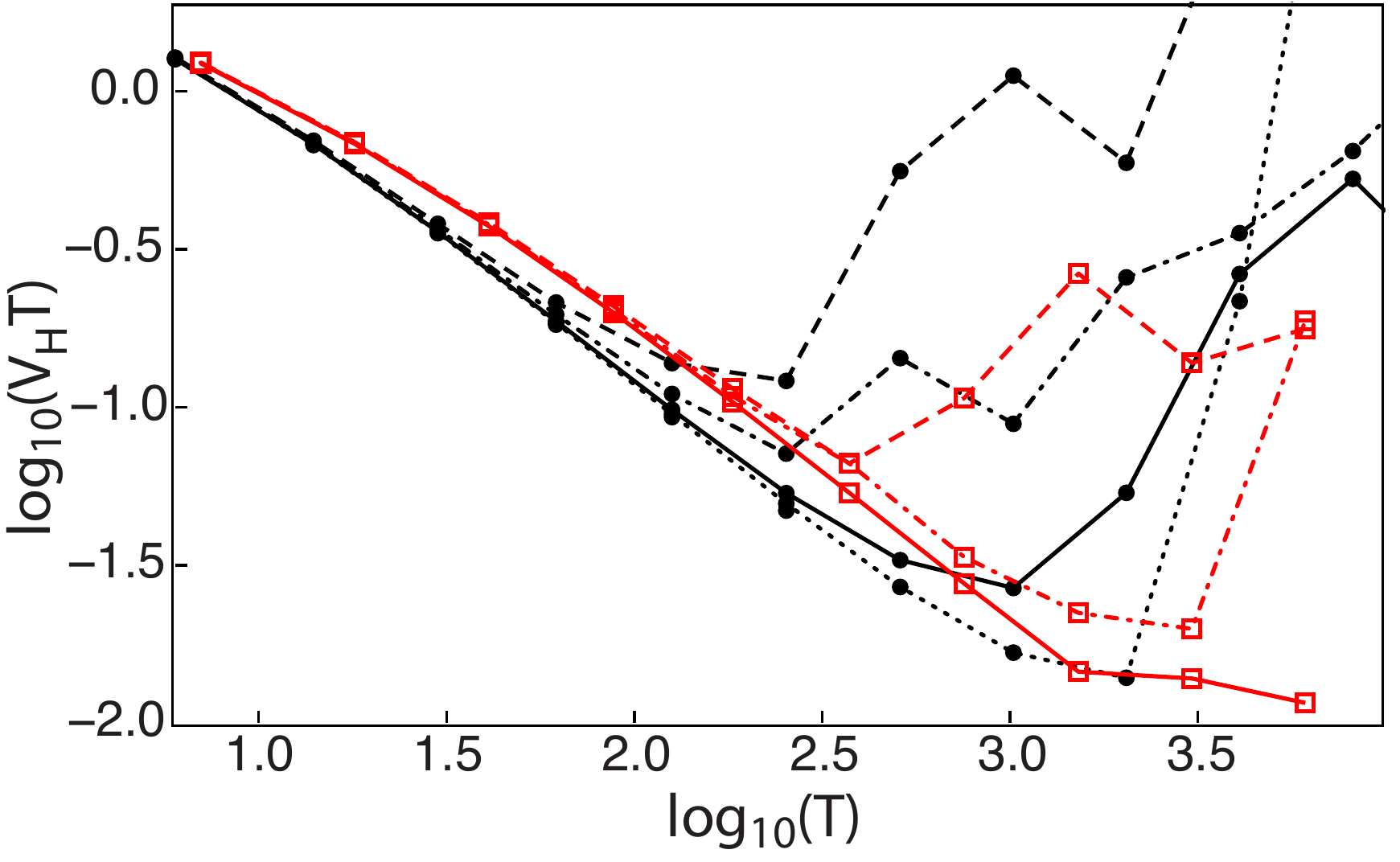}\vspace{-10pt}
\end{center}
\caption{(color online) Holevo variance vs. total time in the presence of signal decay. Black lines with circles: $M=2$ measurements per step. Red lines with squares: M=$n+1$ measurements at the $n^{th}$ step. The relaxation constant $T_2$ was taken to be $T_2/\tau=5\times10^{-4}$ (dotted line), $10^{-3}$ (solid lines), $2.5\times10^{-3}$ (dash-dotted lines) and   $5\times10^{-3}$ (dashed lines). The total time $T$ was measured in units of the dimensionless time $\tau$.}
\end{wrapfigure}

In the main text we discussed the effects on the adaptive scheme of imperfect readout of the sensor state. A similar effect is expected as well if the signal decays due to decoherence during the Ramsey interrogation time, as the difference in the probability of getting a different result ($m=0,1$) given a different phase is reduced by a factor $e^{-\tau/T_2}$:
\[\mathcal{P}_\theta(1|b)-\mathcal{P}_\theta(0|b)=\half e^{-\tau/T_2}\cos(b\tau+\theta).\]
The time-dependence of such an imperfection, though, complicates the recursive relationship (Eq.~3 of the main text), thus we analyze the effect of decoherence numerically. We find that the decay effectively sets a maximum number of measurements (see figure), since the interrogation time cannot 	exceed $T_2$. A small improvement is achieved by increasing the number of measurements per step, but the QML is not recovered at longer times.

Another source of imperfection would derive from variations of the phase during the total estimation time. The rate of this variations  sets an upper limit to the number of steps in the adaptive scheme.

\twocolumngrid

\bibliographystyle{apsrev4P}
\bibliography{../../Biblio}

\end{document}